\documentclass[prb,a4paper]{revtex4}
\usepackage[pdftex]{graphicx}
\usepackage{multirow}

\newcommand{\nc}{\newcommand}
\nc{\be}[1]{\begin{equation}\mbox{$\label{#1}$}}
\nc{\bea}[1]{\begin{eqnarray} \mbox{$\label{#1}$}}
\nc{\Section}[2]{\section{#2}\label{#1}}
\nc{\Bibitem}[1]{\bibitem{#1}}
\nc{\Label}[1]{\label{#1}}

\nc{\eea}{\end{eqnarray}}
\nc{\ee}{\end{equation}}

\nc{\bdm}{\begin{displaymath}}
\nc{\edm}{\end{displaymath}}
\nc{\dpsty}{\displaystyle}
\nc{\bc}{\begin{center}}
\nc{\ec}{\end{center}}
\nc{\ba}{\begin{array}}
\nc{\ea}{\end{array}}
\nc{\bab}{\begin{abstract}}
\nc{\eab}{\end{abstract}}
\nc{\btab}{\begin{tabular}}
\nc{\etab}{\end{tabular}}
\nc{\bit}{\begin{itemize}}
\nc{\eit}{\end{itemize}}
\nc{\ben}{\begin{enumerate}}
\nc{\een}{\end{enumerate}}
\nc{\bfig}{\begin{figure}}
\nc{\efig}{\end{figure}}

\nc{\arreq}{&\!=\!&}
\nc{\arrmi}{&\!-\!&}
\nc{\arrpl}{&\!+\!&}
\nc{\arrap}{&\!\!\!\approx\!\!\!&}
\nc{\non}{\nonumber}
\nc{\align}{\!\!\!\!\!\!\!\!&&}

\def\lsim{\; \raise0.3ex\hbox{$<$\kern-0.75em
      \raise-1.1ex\hbox{$\sim$}}\; }
\def\gsim{\; \raise0.3ex\hbox{$>$\kern-0.75em
      \raise-1.1ex\hbox{$\sim$}}\; }

\nc{\DOT}{\hspace{-0.08in}{\bf .}\hspace{0.1in}}
\nc{\Laada}{\hbox {$\sqcap$ \kern -1em $\sqcup$}}
\nc\loota{{\scriptstyle\sqcap\kern-0.55em\hbox{$\scriptstyle\sqcup$}}}
\nc\Loota{{\sqcap\kern-0.65em\hbox{$\sqcup$}}}
\nc\laada{\Loota}
\nc{\qed}{\hskip 3em \hbox{\BOX} \vskip 2ex}

\nc{\real}{{\rm I \! R}}
\nc{\Z}{{\sf Z \!\!\! Z}}
\nc{\complex}{{\rm C\!\!\! {\sf I}\,\,}}
\def\bigid{\leavevmode\hbox{\small1\kern-3.8pt\normalsize1}}
\def\id{\leavevmode\hbox{\small1\kern-3.3pt\normalsize1}}
\nc{\slask}{\!\!\!/}
\nc{\bis}{{\prime\prime}}
\nc{\pa}{\partial}
\nc{\na}{\nabla}
\nc{\ra}{\rangle}
\nc{\la}{\langle}
\nc{\goto}{\rightarrow}
\nc{\swap}{\leftrightarrow}

\nc{\EE}[1]{ \mbox{$\cdot10^{#1}$} }
\nc{\abs}[1]{\left|#1\right|}
\nc{\at}[2]{\left.#1\right|_{#2}}
\nc{\norm}[1]{\|#1\|}
\nc{\abscut}[2]{\Abs{#1}_{\scriptscriptstyle#2}}
\nc{\vek}[1]{{\rm\bf #1}}
\nc{\integral}[2]{\int\limits_{#1}^{#2}}
\nc{\inv}[1]{\frac{1}{#1}}
\nc{\dd}[2]{{{\partial #1}\over{\partial #2}}}
\nc{\ddd}[2]{{{{\partial}^2 #1}\over{\partial {#2}^2}}}
\nc{\dddd}[3]{{{{\partial}^2 #1}\over
    {\partial #2 \partial #3}}}
\nc{\dder}[2]{{{d #1}\over{d #2}}}
\nc{\ddder}[2]{{{d^2 #1}\over{d {#2}^2}}}
\nc{\dddder}[3]{{d^2 #1}\over
    {d #2 d #3}}
\nc{\dx}[1]{d\,^{#1}x}
\nc{\dy}[1]{d\,^{#1}y}
\nc{\dz}[1]{d\,^{#1}z}
\nc{\dl}[1]{\frac{d\,^{#1}l}{(2\pi)^{#1}}}
\nc{\dk}[1]{\frac{d\,^{#1}k}{(2\pi)^{#1}}}
\nc{\dq}[1]{\frac{d\,^{#1}q}{(2\pi)^{#1}}}

\nc{\bfT}{{\bf T }}

\nc{\cA}{{\cal A}}
\nc{\cB}{{\cal B}}
\nc{\cD}{{\cal D}}
\nc{\cE}{{\cal E}}
\nc{\cG}{{\cal G}}
\nc{\cH}{{\cal H}}
\nc{\cL}{{\cal L}}
\nc{\cO}{{\cal O}}
\nc{\cT}{{\cal T}}
\nc{\cN}{{\cal N}}
\nc{\cR}{{\cal R}}
\nc{\rvac}[1]{|{\cal O}#1\rangle}
\nc{\lvac}[1]{\langle{\cal O}#1|}
\nc{\rvacb}[1]{|{\cal O}_\beta #1\rangle}
\nc{\lvacb}[1]{\langle{\cal O}_\beta #1 |}
\nc{\bb}{\bar{\beta}}
\nc{\bt}{\tilde{\beta}}
\nc{\ctH}{\tilde{\cal H}}
\nc{\chH}{\hat{\cal H}}
\nc{\1}{\aa}
\nc{\2}{\"{a}}
\nc{\3}{\"{o}}
\nc{\4}{\AA}
\nc{\5}{\"{A}}
\nc{\6}{\"{O}}
\nc{\al}{\alpha}
\nc{\g}{\gamma}
\nc{\Del}{\Delta}
\nc{\e}{\textrm{e}}
\nc{\eps}{\epsilon}
\nc{\lam}{\lambda}
\nc{\Om}{\Omega}
\nc{\ve}{\varepsilon}
\nc{\mn}{{\mu\nu}}
\nc{\vp}{\varphi}

\nc{\rf}[1]{(\ref{#1})}
\nc{\nn}{\nonumber \\*}
\nc{\bfB}{\bf{B}}
\nc{\bfv}{\bf{v}}
\nc{\bfx}{\bf{x}}
\nc{\bfy}{\bf{y}}
\nc{\vx}{\vec{x}}
\nc{\vy}{\vec{y}}
\nc{\oB}{\overline{B}}
\nc{\oI}{\overline{I}}
\nc{\oR}{\overline{R}}
\nc{\rar}{\rightarrow}
\nc{\ti}{\times}
\nc{\slsh}{\hskip-5pt/}
\nc{\sm}{Standard~Model~}
\nc{\MP}{M_{\rm Pl}}
\nc{\mpl}{M_{\rm Pl}}
\nc{\tp}{t_{\rm Pl}}

\nc{\pmin}{p_{\rm min}}
\nc{\pmax}{p_{\rm max}}
\nc{\fo}{f_0}
\nc{\foi}{f_{0,i}\,}
\nc{\fop}{f_0^P}
\nc{\fou}{f_0^U}

\nc{\eff}{{\rm eff}}
\nc{\MT}{M_{\rm T}}
\nc{\ML}{M_{\rm L}}
\nc{\kk}{\vek{k}}
\nc{\pp}{{\rm p}}
\nc{\pt}{\partial_t}
\nc{\half}{{1\over 2}}
\nc{\w}{\omega}
\nc{\uhat}{\hat{U}_\w}

\nc{\etal}{\mbox{\it et al.}}
\nc{\ie}{{\it i.e. }}
\nc{\eg}{{\it e.g. }}
\nc{\trh}{T_{\rm RH}}
\nc{\ad}{{a'\over a}}
\nc{\bd}{{b'\over b}}
\nc{\Rd}{{R'\over R}}
\nc{\diag}{{\textrm{diag}}}
\nc{\mato}[1]{\tilde{#1}}
\nc{\sech}{\textrm{sech}}
\nc{\I}{\textrm{I}}
\nc{\II}{\textrm{II}}
\nc{\III}{\textrm{III}}
\nc{\vev}[1]{\langle #1 \rangle}
\nc{\hyp}{\,\; F_{1{\hskip -16pt}2}{\hskip 11pt}}
\nc{\brhom}{\overline{\rho}_M}
\nc{\brho}{\overline{\rho}}
\nc{\rhob}{\overline{\rho}}
\nc{\Pb}{\overline{P}}
\nc{\bH}{\overline{H}}
\nc{\ep}{{1+4\eps}}

\nc{\lcdm}{$\Lambda$CDM}
\nc{\ms}{\langle\sigma\rangle}

\def\smiley{\hbox{\large$\bigcirc$\hspace{-.80em}%
\raise.2ex\hbox{$\cdot\cdot$}\kern-.61em    
\lower.2ex\hbox{\scriptsize$\smile$}}\ }

\def\frowney{\hbox{\large$\bigcirc$\hspace{-.80em}%
\raise.2ex\hbox{$\cdot\cdot$}\kern-.635em
\lower.2ex\hbox{\scriptsize$\frown$}}\ }

\begin{document}

\title{\bf Stability of non-homegeneous models and fine tuning of initial state}

\author{P.~Sundell}
\author{I.~Vilja}
\affiliation{Department of Physics and Astronomy, University of Turku, 
FI-20014 Turku, Finland}
\date{\today}

\begin{abstract}
We apply phase space analysis to inhomogeneous cosmological model given by
Lema\^{i}tre-Tolman model. We describe some general conditions required to interpret
the model stable enough and, in the present paper, apply them to two special cases: 
dust filled homogeneous model with and without cosmological constant. We find that
such stability explaining all present astrophysical observations can not be achieved
due to instabilities in phase space. This hints that non-homogeneous models are not likely
to be physically viable, although any conclusive analysis requires more realistic modeling of 
non-homogeneous universe. 
\end{abstract}

\maketitle

\section{Introduction}

Supernovae observations \cite{Riess1998, Perlmutter1999} made just 
before the break of the millennium implies that the universe appears to be expanding 
at an increasing rate. A bit later made observations of cosmic microwave background 
(CMB) radiation \cite{Spergel} supports the conclusions made out of the supernovae 
observations. The most popular ways of explaining these observations are with models based on
Friedmann-Lema\^{i}tre-Robertson-Walker (FLRW) metric, which is based on general 
relativity and the principles of cosmological and Copernican. The cosmological 
principle merely states that the universe is spatially homogeneous everywhere, 
whereas the Copernican principle states that there are no preferred points in the 
universe.  FLRW  metric and models based on it are extensively presented in the literature of cosmology 
(see for example \cite{KolbTurner1994}).

Even though the FLRW based models fit well inside the frame provided by our observations, 
it is also for long known to suffer some problems, for example the fine tuning
problem \cite{KolbTurner1994, Weinberg2008} and the cosmological constant problem 
\cite{KolbTurner1994, Weinberg2000}. One of the strengths of the FLRW based models 
is  simplicity due to varies approximations, although, that can also be counted as a
weakness. It is also questionable if all approximations are made acceptably, as is 
pointed out by Shirokov and Fisher \cite{ShirokovFisher1963}, where is questioned 
if the homogeneity approximation should be done to the Einstein tensor 
$G_{\mu \nu}$, rather than to the metric $g_{\mu \nu}$, since in general 
$<G_{\mu \nu}(g_{\mu \nu})>\neq G_{\mu \nu}(<g_{\mu \nu}>)$. However, the problem is 
more complex than this. As pointed out by Shirokov and Fisher 
\cite{ShirokovFisher1963}, Einstein equations are no longer tensor equations after 
averaging in the sense, that they can not be changed {\it e.g.} from covariant form to 
contravariant form with metric tensor without altering the equations. In this 
perspective it seems, that only tensors rank 0 and scalars have well validated 
averages. For this kind of approach see Buchert \cite{Buchert1999}, where he 
transforms the Einstein equations into scalar equations before averaging.
It is very much possible that the problems  FLRW based models suffer are due to 
approximations, and solely the work of Shirokov, Fisher and Buchert implies that 
the first approximation to investigate more is the homogeneity. Such a model was first 
introduced by Lema\^{i}tre \cite{Lemaitre}, and later on studied by Tolman \cite{Tolman} 
and its called Lema\^{i}tre-Tolman (LT) model. For 
further developments of the model see \cite{Bondi, BKHC2010}.
The LT model have not yet been studied as widely as  FLRW  based models, hence all the 
problems it suffers have probably not yet been discovered, but it have already shown 
its power by overcoming some of the problems of the FLRW based models. For example, Mattsson 
has shown that the LT model can explain the main cosmological observations without 
dark energy.\cite{Mattsson} 

Actually, FLRW based models are often presented including early times inflation, which solves 
the fine tuning problem. This solution can not be generalized to LT model {\it a prior}, because 
the evolution of the universe in LT model is dependent in coordinate distance, which would make 
the inflation occur differently in separate locations, and the consequences of this are unknown. 
However, inflation is not the only possible explanation for the homogeneous tendency of cosmological 
observations. We explore the possibility that the structure of the equations governing the evolution 
of the universe is such, that it has a inbuilt property to make everything appear as observed. We 
study the existence of this property by using phase 
space analysis. The aim is to find restrictions to viable and stable solutions of 
the differential equations governing the universe in the LT model. Viable solutions 
mean here that the solutions are  consistent with the observations. Stable 
solutions include such solutions, which are attracted towards the universe 
 we observe; therefore these kind of solutions do not need fine tuning of initial state. Especially 
interesting cases are where the dark energy is absent.

It is also interesting to see, if our results offers insight to the homogeneity 
approximation. By that we mean, if all the viable and stable LT models are 
approximately homogeneous. This subject however is not going to be important in this 
paper, but it is merely pointed out as a possibility what more can our results 
offer.

The focus of the present paper is to introduce a novel method to use stability analysis and its 
general features with the Lema\^{i}tre-Tolman model, which is done in section II. In section IIA, homogeneous 
cases in general are applied to the methods found out. In sections IIB and IIC pressure free, flat and 
homogeneous universe is investigated, in the cases of dust filled and dust and dark energy filled universes, using the phase space analysis. Finally in section III the results are discussed. Realistic applications where 
viable and stable inhomogeneous models are to be determined are left to forthcoming publications.

\section{Lema\^{i}tre-Tolman model and phase space analysis}

The LT model\cite{BKHC2010} describes a dust filled inhomegeneous 
but isotropic universe which energy momentum tensor reads as 
$T^{\mu \nu}=\rho u^{\mu} u^{\nu}$.\footnote{We consider the special case of the model where the observer is at the origin.} Here $\rho=\rho(t,r)$ 
is matter density and $u^{\mu}$ is the local four-velocity. As the coordinates are 
assumed to be comoving, the four-velocity is simply $u^{\mu}=\delta^{\mu}_t$. 
The standard synchronous gauge metric in the LT model is given 
by\footnote{We use units in which $c=1$.}
\be {LT_metric}
ds^2=-dt^2+\frac{R_r^2 dr^2}{1+f(r)}+R(t,r)^2 \left( d \theta^2 +
\sin^2 \theta\, d\phi^2 \right),
\ee
where the subscript $r$ denotes derivative with respect to radial coordinate, 
$R_r=\partial R(t,r)/\partial r$, and $f(r)> -1$. In this prescription the 
evolution of universe is built in to the local scale factor $R$ whereas function $f$
controls the overall radial dilatation. Including the cosmological constant $\Lambda$
into the Einstein equations, after some integrations, one obtains the relevant
differential equations as 
\be {GFE}
\dot{R}^2=\frac{2M}{R}+f+\frac{\Lambda}{3}R^2,
\ee 
and 
\be {DEQ}
\kappa \rho=\frac{2M_r}{R_r\,R^2},
\ee
where $M=M(r)$ is a arbitrary function, $\kappa=8\pi G/c^4$, and $G$ is the 
Newton's constant of gravity. 

In general, all the quantities (or functions) in Eqs. (\ref{GFE}) and (\ref{DEQ}) are unknown. 
However, if they are presented as quantities dependent only on redshift, they can be received 
from cosmological observations, or be derived from the obsevations, {\it e.g.} if $R(z)$ is 
known we get $R_z(z)$ as its derivative. In the LT models   the redshift equation reads as 
\be  {dz/dr}
\frac{dz}{dr}=\frac{(1+z)R_{rt}(r,t)}{\sqrt{1+f}},
\ee 
and the angular diameter distance $d_a$ is  $d_a(z)=R(t(z),r(z))$.
\cite{MustaphaHellabyEllis1998,MustaphaBassettHellabyEllis1998,Mattsson} 

The relation between $t$ and $r$ can be given as follows. The path of radial light ray is given by 
the radial null geodesic, where $ds^2=d\theta^2=d\psi^2=0$. Using metric (\ref{LT_metric}) it is
\be {null_geodesic}
dt=\pm \frac{R_r }{\sqrt{1+f}}dr,
\ee
where the signs correspond an incoming ($-$) and  an outgoing ($+$) light rays. 
It is possible\cite{LuHellaby2007} to choose radial coordinate $r$, {\it i.e.} use remaining gauge freedom so, 
that for incoming ray $dr= -dt$. So, along the ray 
\be {t_wrt_r}
t=t(r)=t_0-r,
\ee
where  $t=t_0$ refers to present time.
For any quantity, a hat $\ \hat{}\ $ denotes that it is evaluated along (incoming) 
light ray, 
{\it e.g.} $\hat R=R(t(r),r)$. Thus, the gauge condition for Eq. (\ref{t_wrt_r}) for 
incoming light can be written as
\be {null_geodesic_gauge}
1= \frac{\hat{R_r} }{\sqrt{1+f}}\, .
\ee

On the radial null geodesic it can be shown\cite{LuHellaby2007} that the relation between the 
matter density $\rho(t,r)$ and the number density of light sources in redshift 
distance\footnote{The number density of sources in redshift distance means the amount of 
sources per steradian per unit redshift interval.} $n=n(z)$ 
is given by
\be {rho_hat}
\hat{\rho}=\frac{\mu n}{\hat{R}^2}\frac{dz}{dr},
\ee
where $\mu=\mu(z)$ is the mean mass per source at given redshift distance.  After some manipulation, 
using Eqs. (\ref{GFE}), (\ref{DEQ}), (\ref{t_wrt_r}), (\ref{null_geodesic_gauge}), and (\ref{rho_hat})
the redshift equation (\ref{dz/dr}) can be cast in the form\cite{LuHellaby2007}
\be {DE}
J(z) z_r+K(z) z_r^2+L(z)z_{rr}=0,
\ee
where we have defined\footnote{Note that here  
$\hat{\rho}=\frac{\mu n z_r}{\hat{R}^2}$ is already substituted 
into the null Raychaudhury equation.}
\bea {J,K,L}
J(z)&=& \kappa (1+z)  \mu n, \nonumber \\
K(z)&=& 2 R \left[ R_z  +  (1+z)   R_{zz}\right], \\
L(z)&=& 2(1+z) R R_z. \nonumber
\eea
All the quantities in Equations $J$, $K$, and $L$ are only dependent on redshift $z$, and they are 
all measurable or they can be derived from measurable quantities. Moreover, because it is evident that 
the quantities dependent only on redshift are evaluated along  light ray, every function or quantity dependent only on $z$ is presented without a 
hat.  Especially now $R=R(z)=\hat{R}(t(r(z)),r(z))$, and hence 
$R_z=\partial R(z)/\partial z= d R(z)/dz$ and $R_{zz}=\partial^2 R(z)/\partial z^2= d^2 R(z)/dz^2$.
Eq. (\ref{DE}) is a second order non-linear differential equation, from which can be solved $z$ 
as a function of $r$. However, that is not the interest here, but rather investigating the 
stability of different solutions. We use phase space analysis\cite{Strogatz1994, Percival1982} for investigation.
Here is chosen   $z_r=y$ to give
\be {psa_general}
\left\{ \begin{array}{l l} 
y_r&=-y\frac{J(z) + K(z)  y}{L(z)} \\ 
z_r&=y .
\end{array} \right.
\ee
Note, that we have given the redshift in terms determined comoving coordinate $r$, which is 
related along the ray to time as $r=t_0-t$. This means that we can use practically
interchangeably the two parameter $r$ or $t$. 

Analyzing cosmological data will give us functions $J$, $K$, and $L$ explicitly with respect to $z$, 
but even now when the explicit forms are unknown, general remarks can be done on what kind of systems Eq. (\ref{DE}) 
can describe. The phase plane can be divided into sections each having its own characteristics. 
Some of the properties can be read out from general properties of the equations and represented 
as a flow plot in $zy$-plane.

The $y$-axis represents location of observations (Earth) at present time; the redshift is there zero. 
On the right 
hand side of the $y$-axis is past or distant objects and the left hand side of the 
$y$-axis can be interpreted as future. On the $z$-axis $z_r=z_t=0$, hence it 
represents apparently static universe. Above the $z$-axis $z_r>0$, the area represents 
apparently (and locally) expanding space, and below the $z$-axis $z_r<0$ the apparently contracting one. 
Therefore, due to the observations, we are mostly interested of the first quarter 
of a phase space plane, past of an expanding universe; however, we do not want to exclude other 
parts of the phase plane {\it per se}, but we give the first quarter of the phase space plane most of our interest.

\begin{figure}[h]
\centering
\includegraphics[scale=0.6]{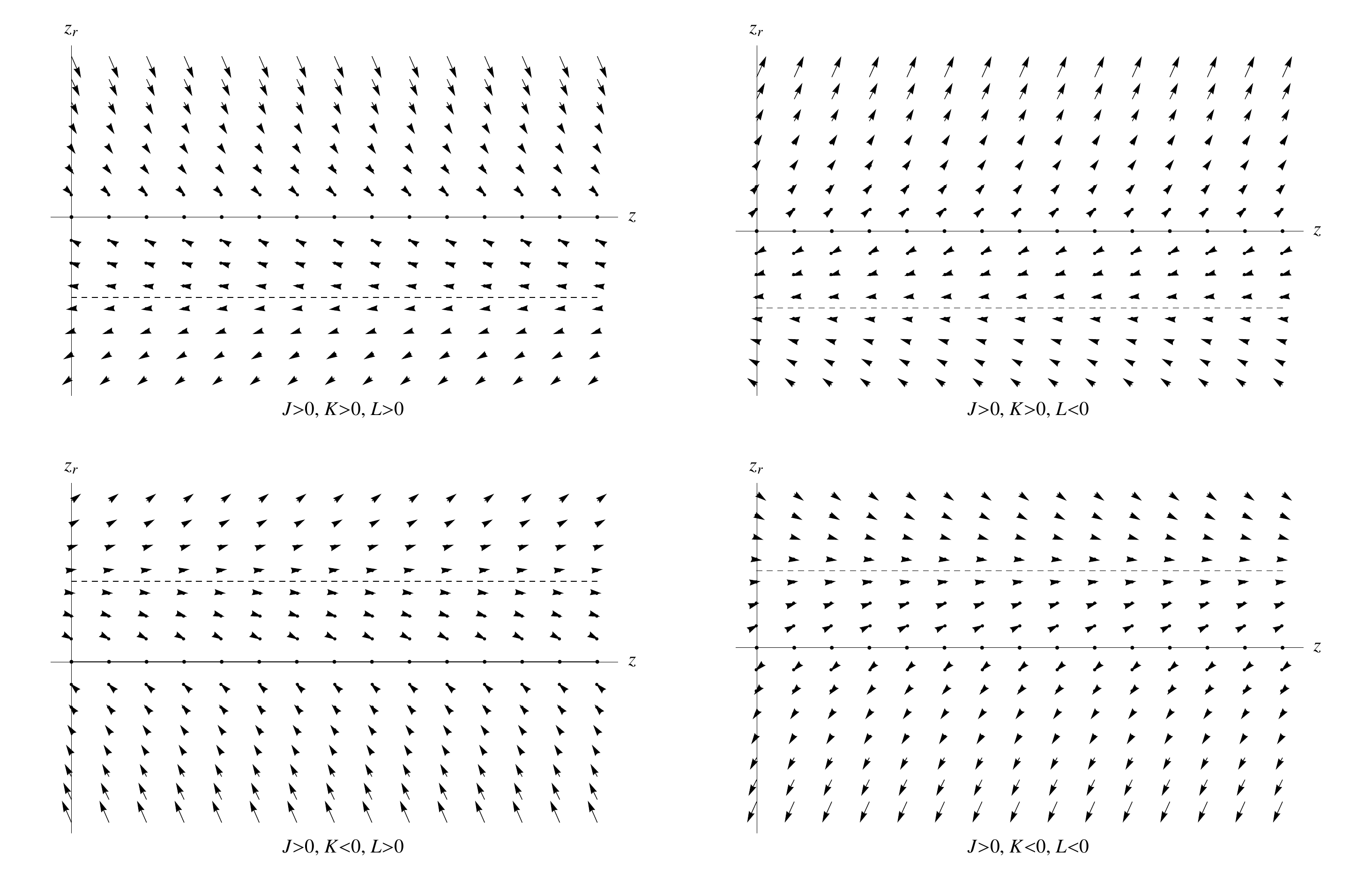} 
\caption{\footnotesize Four different situations depending on the signs of $K$ and $L$ that can occur on a 
phase plane with the system (\ref{psa_general}). Functions $J$, $K$, and $L$ are here chosen to be $+1$ or $-1$ 
for simplicity. The dashed curve is the nullcline $J + K y=0$.} 
\label{k1}
\end{figure}

Looking to the latter of the Eqs. (\ref{psa_general}) one sees, that whenever $y>0$ 
the flow arrows points right, and when $y<0$ they points left. The curves where $y_r$ or $z_r$ is zero are 
called nullclines and the points where $y_r$ and $z_r$ is zero are called fixed points. In the system 
(\ref{psa_general}) all the points on 
curve $y=0$ are fixed points, which means that the curve $y=0$ is fixed. The curves where $L(z)=0$ needs special 
attention as the system (\ref{psa_general}) is not well defined there. 
The physical perspective also gives more restrictions. The form of function $J$ 
reveals that it is positive implying that the curve $J(z) + K(z) y=0$ 
can not cross the $z$-axis. The only exceptions are at the origin of the spherical symmetry ($r=0$), where 
it is required to have $R(t,0)=0$ for all $t$ to avoid point mass and curvature singularity at $r=0$, and 
possibly at the Big Bang or the Big Crunch.\cite{BKHC2010} Requirement $R(t,0)=0$ in gauge (\ref{t_wrt_r}) 
is $R(t_0,0)=0$, which is compatible with $R(z)$ being zero at $z=0$ as $R(z)$ is the angular diameter distance. 
Because $J$ is always positive (neglecting the special points discussed above) and $K$ and $L$ seem to able 
to change their signs, combining these functions with different allowed signs there is essentially four 
different types of situations (in the case of system (\ref{psa_general})) that can appear on the phase 
portraits, presented in Fiq. \ref{k1}. The nature of the function $R(z)$ implies that $L$ is zero 
only if $R_z(z)$ is zero (neglecting again the special points discussed above) and because $R_z(z)=R_r(r)r_z$, 
is then either $R_r(r)$ or $r_z$ zero. The latter case can approach to zero when $y \rightarrow \infty$, and 
$R_r(r)=0$ occurs at the apparent horizon\cite{LuHellaby2007}.

Consider now the situation where functions $J$, $K$, and $L$ are formed from observational data. 
Eq. (\ref{psa_general}) is now given explicitly, so phase plane can be drawn and the best fit curve 
compared to observations can be fitted. It is assumable that the best fit curve coincides closely 
to the best fit to the observations in the isotropic and homogeneous models, which means that the 
best fit curve is not very "lumpy"  and is approximately monotonic. Now, consider each observable 
separately. Each of them have their own trajectory, which are unknown to us, because we do not have 
enough data from each observable. But if we had enough data from each observable, we could draw their 
trajectories on the same phase plane that the best fit curve is on. So, each observable have their 
own trajectory, but still they seem to sit on a approximately monotonic curve, the best fit curve. 
In our scheme this appears to happen only, if the flow arrows are pointing towards the best fit curve. 
This would ensure that for large number of initial values, observables end up (after long enough 
time period) near by to the best fit curve. This issue, however, is not that simple and it will be 
discussed more in  section  Conclusions and discussion.

At simplest the interest in phase plane analysis is concentrated into the existence and
properties of the fixed points. In this case however, observations suggests that the best fit curve 
should be attractive. In the system (\ref{psa_general}) only $z$-axis can be referred as a attractive 
curve, but its nature as apparent static universe is not what is observed. 
However, nullclines can be thought of to be attractive like, since flow arrows can point towards them 
as in {\it e.g.} the low-right case in Fig. \ref{k1}. In our system the only physically interesting 
possible attractive like  nullcline is $J(z) + K(z) y=0$. However, the exact  identification between the 
nullcline and the best fit curve can not be done, because then both Eqs. $J+Ky=0$ and (\ref{DE}) should 
be satisfied simultaneously causing either the solution $y(z)$ be linear or $L(z)=0$, but it is not even 
necessary as long as the identification can satisfy observations inside their inaccuracy limits. 

The properties of the system (\ref{psa_general}) discussed above implies that the curve $J+Ky=0$ can be 
attractive in the first quarter of the phase plane only if $J$ is positive and $K$ and $L $ are negative. Even though these boundaries are 
necessary, they are far from sufficient for the following reason. Consider a situations where $0<J$, and 
$K,L<0$ and the value of the slope of the curve $J+Ky=0$, $ -J/K $, is positive. On the curve $J+Ky=0$ is 
always $y_r=0$, thus flow arrows on the curve are horizontal and pointing right, {\it i.e.}, the slopes of 
the flow arrows $y_z=y_r/z_r$ are zero. Now, approaching the curve from below vertically makes $y_z$ to approach 
zero. At some distance to $J+Ky=0$ before $y_z$ reaches to zero, it becomes smaller than $-J/K$. This means 
that the solution is no longer approaching the curve $ J+Ky=0$, because flow arrows determine the slopes and 
the progressing directions of the solutions in each point. Similar situation occur, if at fixed $z$ we approach 
the curve $J+Ky=0$ from above and $-J/K<0$. This is also why nullclines are rather attractive like than attractive.
As it follows, we need more sufficient methods to measure the attractivity of the nullcline $J+Ky=0$. Especially, 
we need a method to measure the distance from the nullcline where it stops acting as an attractor. In the present 
paper for this is used the following method.

Let us consider our system on area $z,y>0$ with positive $J$ and negative $K$ and $L$. The nullcline $J(z) + K(z) y=0$ 
is attractive like, if flow arrows around it points towards it. Now below the nullcline flow arrows point up-right and 
above the curve they point down-right. 
The slope of the nullcline is positive and all the flow arrows above it points towards it, but the flow arrows below 
it points towards it only if the slopes of the flow arrows $y_z(z)$ at some given $z$ are greater than the slope of 
the nullcline $-J/K$ at the same $z$. 
Let $\delta_y>0$ be the (vertical) distance from the nullcline $J(z) + K(z) y=0$. The slope of the flow arrow at 
given $z$ and distance $\delta_y$ below the nullcline is:

\bea {eq1}
y_z^-& := &\left[\frac{(y-\delta_y)_r}{z_r}\right]_{y=-J(z)/K(z)}\nonumber\\
&=&\left[ \frac{-(y-\delta_y)\frac{J(z) + 
K(z)  (y+\delta_y)}{L(z)}}{y-\delta_y}\right]_{y=-J(z)/K(z)}\nonumber \\
&=&-\left[\frac{J(z) + K(z)  (y-\delta_y)}{L(z)}\right]_{y=-J(z)/K(z)} \nonumber\\
&=& \frac{ K(z)  \delta_y}{L(z)}-\left[\frac{J(z) + 
K(z)  y}{L(z)}\right]_{y=-J(z)/K(z)}  \nonumber\\
&=& \frac{ K(z) }{L(z)} \delta_y.
\eea 
From Eq. (\ref{eq1})
one can explicitly see, that the size of the vertical step  $\delta_y$ taken from the nullcline  effects to the 
attractiveness in a linear fashion: the larger the step is, more attracitve like the nullcline seems. This means 
that $\delta_y$ takes a role of a parameter comparable to observational inaccuracies.
Hence, the slopes of the flow arrows $y_z(z)$ are greater than the slope of the nullcline $-J/K$ at at some 
given $z$, if
\be{eq3}
\frac{ K(z) }{L(z)}\delta_y> y_z = -\frac{J_z(z)}{K(z)}+\frac{J(z)}{K(z)^2}K_z(z).
\ee
The above inequality is the restriction we use to study the attractiveness of the nullcline $J+Ky=0 $.

\subsection{Isotropic and homogeneous models}

To give more concrete touch of our prescription, we check how the method 
works with isotropic and homogeneous models. Note, that FLRW metric can not be 
used here, because it is not compatible with our gauge choice $dt=-dr$. This can be explicitly
seen by comparing the standard  FLRW metric relation between $r$ and $z$ given by\cite{KolbTurner1994}
\be {r(z)_in_RW_1}
r=\frac{2 z \Omega _0+(2 \Omega _0-4) \left(\sqrt{z \Omega _0+1}-1\right)}{a_0 H_0 
(z+1) \Omega
   _0^2},
\ee 
with $a=a_0/(1+z)$ to the relation derived from Eq. (\ref{null_geodesic_gauge}),
$a=\sqrt{1-k r^2}$  giving
\be {r(z)_in_RW_2}
r=\pm \sqrt{\frac{1}{k}-\frac{a_0^2}{k(1+z)^2}}.
\ee
Clearly Eqs. (\ref{r(z)_in_RW_1}) and (\ref{r(z)_in_RW_2}) are not equivalent, 
except in some special cases. However, it is necessary to write isotropic and homogenous
space-time in Robertson-Walker coordinates, and we can proceed by other means. 

In spherically symmetric homogeneous space-time 
the metric can always be given as\cite{Weinberg1972}
\be {eq4}
ds^2=g(v) dv^2+f(v)\left(d\textbf{u}^2+\frac{k(\textbf{u}\cdot d 
\textbf{u})^2}{1-k\textbf{u}^2} \right),
\ee
where $v$ and $\textbf{u}=(u_1, u_2, u_3)$ are the coordinates, $g(v)$ is a 
negative and $f(v)$ is a positive function of $v$, and $k$ is spatial curvature 
and can be chosen to be 1, 0, or -1. For our purposes it is convenient to define 
new coordinates $t$, $r$, $\theta$, and $\varphi$ by
\bea {eq5}
 \sqrt{-g(v)}dv&=&dt, \nonumber\\
u_1&=&\Theta (r) \sin \theta \cos \varphi, \nonumber\\
u_2&=&\Theta (r) \sin \theta \sin \varphi, \nonumber\\
u_3&=&\Theta (r) \cos \theta. 
\eea
Then we have
\be {IH_metric}
ds^2=-dt^2+a^2(t)\left[ \frac{\Theta_r^2}{1-k \Theta^2}dr^2+
\Theta^2 d\Omega^2 \right],
\ee 
where $a(t)=\sqrt{f(v)}$ and $d\Omega^2=d\theta^2+\sin^2\theta d\varphi^2$.
With the above metric definition the field equations take the normal Friedmann form and
 therefore can be written as
\be {a_t_general}
a_t=H_0 a \sqrt{\sum_i \Omega_i^{(0)} (a/a_0)^{-3(1+w_i)} },
\ee
where $\Omega_i^{(0)}$ is the present value of the energy density of different energy forms.
For dust, dark energy, and spatial curvature, the $w_i$ takes values $0,$ $-1$, and $-1/3$ respectively. 
Because the LT model does not take pressure into account, for our purposes it is unnecessary to include relativistic matter here either, even though generally it could be done.

To be able to use the Eqs. (\ref{psa_general}),
we need to specify quantities $\mu$, $n$ and $R$. 
According to Eq. (\ref{rho_hat}), $\mu$ and $n$ can be given with $R$, $\rho$, and $z_r$,
and because 
in metric (\ref{IH_metric})  $R=a\Theta$, we need to find out presentations for 
quantities $a$, $\Theta$, $\rho$, and $z_r$ with respect to $z$.

In LT model, the energy momentum tensor $T^{\mu \nu}$ includes only dust, hence 
\begin{equation}
\rho = \rho^{(0)} (a_0/a)^3,
\end{equation}
where  $\rho^{(0)}= \Omega_M^{(0)}\frac{3H_0^2}{\kappa}$, and $\Omega_M^{(0)}$ is the present dust 
(or non-relativistic matter) density of the universe.

With metric (\ref{IH_metric})  equations (\ref{null_geodesic_gauge}) and 
(\ref{dz/dr}) reduces to be
\be {null_geodesic_gauge_HI}
1= \frac{a \Theta_r }{\sqrt{1-k\Theta^2}}
\ee 
and
\bea{dz/dr_HI}
\frac{dz}{dr}=\frac{(1+z)a_t \Theta_r}{\sqrt{1-k\Theta^2}} .
\eea
Combining these the usual Robertson-Walker relation 
\be{dz/dr_HI2}
\frac{1}{1+z}=\frac{a}{a_0}
\ee
is reproduced and as $r=t_0-t$ is $z_r=a_0 a_t/a^2$, we can write
\be {observations2}
z_r=H_0 (z+1) \sqrt{\sum_i \Omega_i^{(0)} (z+1)^{3(1+w_i)}}.
\ee
In the given gauge, the function 
$\Theta$ can be calculated directly from Eq. (\ref{null_geodesic_gauge_HI}), 
which can also be written as
\be {eq6}
\int_{\Theta_0}^{\Theta}\frac{d\Theta}{\sqrt{1-k \Theta^2}}=\int_0^z \frac{1+z}{a_0}
\frac{1}{z_r}dz,
\ee 
where we have used the relation
\be {eq7}
\int_0^r \frac{dr}{a}=\int_0^z \frac{1}{a}\frac{dr}{dz}dz=\int_0^z \frac{1+z}{a_0}
\frac{1}{z_r}dz\, .
\ee
 $\Theta(r=0)=0$ is determined by the requirement $R(z=0)=0$.
The solution for curvature cases can now be integrated out.
We find
\be{eq9}
\Theta =\sinh \left( \int_0^z \frac{1+z}{a_0}\frac{1}{z_r}dz \right), \quad 
\quad k=-1\,,
\ee
\be{eq10}
\Theta =\int_0^z \frac{1+z}{a_0}\frac{1}{z_r}dz, \qquad \qquad \qquad k=0\,,
\ee
\be{eq11}
\Theta =\sin \left( \int_0^z \frac{1+z}{a_0}\frac{1}{z_r}dz \right), \quad k=1\, .
\ee
In the following sections some numerical calculations are executed. To carry out the numerics we have 
chosen $100\, {\rm km/s/Mpc}=1$, which makes all the physical quantities dimensionless. For subsequent calculations also it is convenient to define a function
\be{g}
g:= \frac{1}{H_0}\left[ -\frac{J_z(z)}{K(z)}+\frac{J(z)}{K(z)^2}K_z(z) \right] \frac{L(z) }{K(z)},
\ee
which in homogeneous models is $g=g(z;\Omega_i^0, a_0, H_0)$, {\it i.e.}, dependent on variable $z$ and parameters $\Omega_i^0$, $a_0$, and $H_0$. Even further, it is easy to see, that in flat homogeneous cases $g=g(z;\Omega_i^0)$, thus the boundary condition (\ref{eq3}) can be written as
\be{eq3h}
\frac{\delta_y}{H_0}>g(z;\Omega_i^0).
\ee
The fact that $g(z;\Omega_i^0)$ do not include parameters $a_0$, $H_0$, or $\kappa$, considerably simplifies the boundary condition.

\begin{figure}[th]
\centering
\includegraphics[scale=0.82]{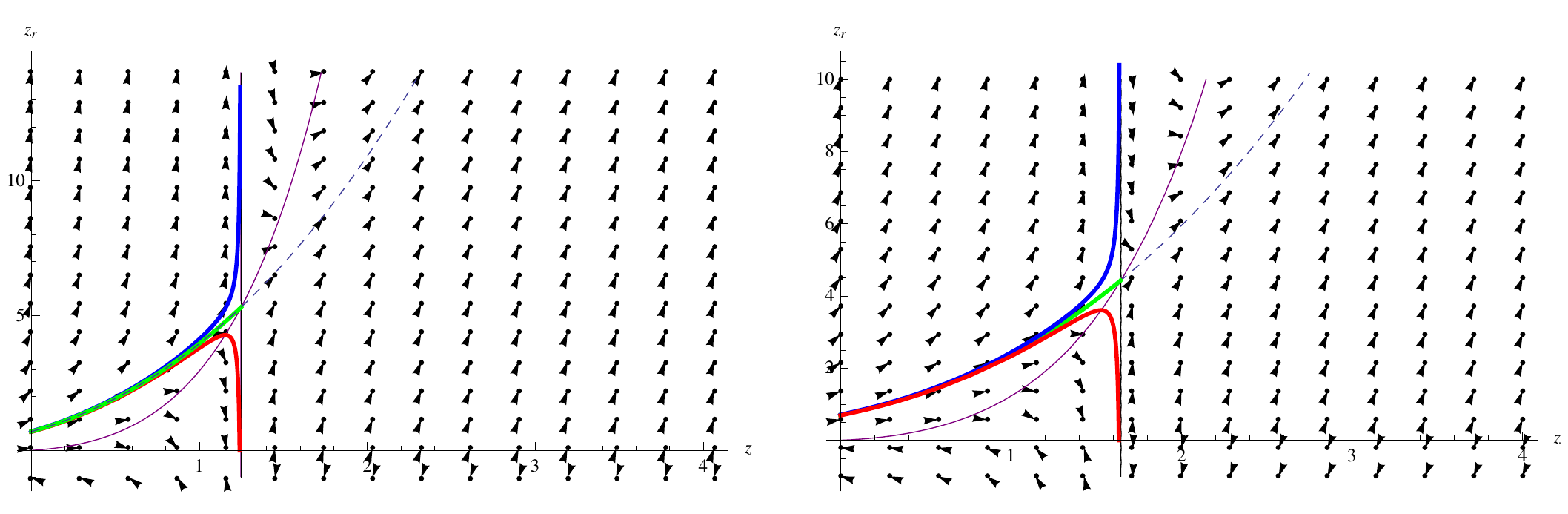} \caption{\footnotesize Phase space portrait of the dust 
filled homogeneous universe (left hand side plane), when $H_0=0.70$, $\Omega_m=1$, and $\Omega_{\Lambda}=0$, and 
of the dust and dark energy filled homogeneous universe (right hand side plane), when 
$H_0=0.700$, $\Omega_m=0.279$, and $\Omega_{\Lambda}=0.721$. In both planes: thin purple curve is $J+Ky=0$, 
thin black curve is $L=0$, and red, green and blue thick curves are solutions of the system (\ref{psa_general}) 
with $H_0$ values 0.69, 0.70, and 0.71 respectively. } 
\label{kuva_molemmat_faasi} 
\end{figure}

\subsection{Dust filled universe}
Let us study stability of the flat dust-filled homogeneous universe, which is often used approximation to investigate the evolution of the universe during matter dominated era.
From Eq. (\ref{observations2}) we obtain redshift relation
\be {matterdonation_z_r}
z_r=H_0 (z+1) \sqrt{ \Omega_m^{(0)} (1+z)^{3}},
\ee
where $\Omega_m^{(0)}=1$ is set from now on.
The pair of differential equations (\ref{psa_general}) 
is now
\bea{eq15}
y_r&=&\frac{y \left(y \left(9 z-4
\sqrt{(z+1)^3}+9\right)-6 H_0 (z+1)^2
\left(z (z+2)-\sqrt{(z+1)^3}+1\right)\right)}{2
(z+1) \left(3 z-2 \sqrt{(z+1)^3}+3\right)}, \\
z_r&=&y, \nonumber
\eea 
which is defined for non-negative $z$ except when $L(z)=0$ at $z=5/4$, which corresponds to $R_z=0$ as 
can be seen from (\ref{J,K,L}). Curve $J+Ky=0$ of the system (\ref{eq15}) can be attractive like on 
the first quarter of the phase plane when $L$ and $K$ are negative and $J$ is positive, which occurs 
at range $5/4<z<65/16$, assuming $H_0>0$. A situation of this kind is illustrated in Fig. 
\ref{kuva_molemmat_faasi}. The case in the figure does not show any signs of attractivity: it seems 
that at the range $5/4<z<65/16$ none of the flow arrows below the curve $J+Ly=0$ would lead any 
solutions towards it. This deduction is strengthen by (\ref{eq3h}), which (by assuming $H_0>0$) in this 
case reduces to be: 
\be{ineq}
g_m(z)=30 \left(\sqrt{\frac{(z+1)^5 \left(7339-8 z \left(64 z^2-516
z+963\right)\right)^2}{(65-16 z)^6}}+\frac{8 (z (112 z-883)+1192) (z+1)^3}{(16
z-65)^3}\right) <\frac{ \delta_y }{ H_0 },
\ee
where the subscript $m$ marks that this is the explicit form of the function $g(z;\Omega_i⁰)$ in the dust filled case. 
At the range $5/4\leq z \leq 65/16$ the function $g_m(z)$ is monotonically increasing from zero to infinity, 
which suggest that the nullcline $J+Ky=0$ of the system (\ref{eq15}) can hardly be attractive like throughout 
the gap $5/4\leq z \leq 65/16$.

The left hand side portrait in Fig. \ref{kuva_molemmat_faasi} illustrates what happens close by $z=5/4$, 
where the system (\ref{eq15}) is not defined. Even a slightest variation from today's observed value of $H_0$ 
seems to lead to a very different kind of universe. 
It is notable, that numerical calculations do not give the same answers for Eqs. (\ref{eq15}) and 
(\ref{matterdonation_z_r}) even with $H_0=0.7$, since the solution of Eqs. (\ref{eq15}) can not cross the 
point $z=5/4$. The area where $J$ is positive, and $L$ and $K$ are negative, does not seem attractive at all, 
and only very close to the singularity it might be attractive like according to $g_m(z)$.

\begin{figure}[th]
\centering
\includegraphics[scale=0.85]{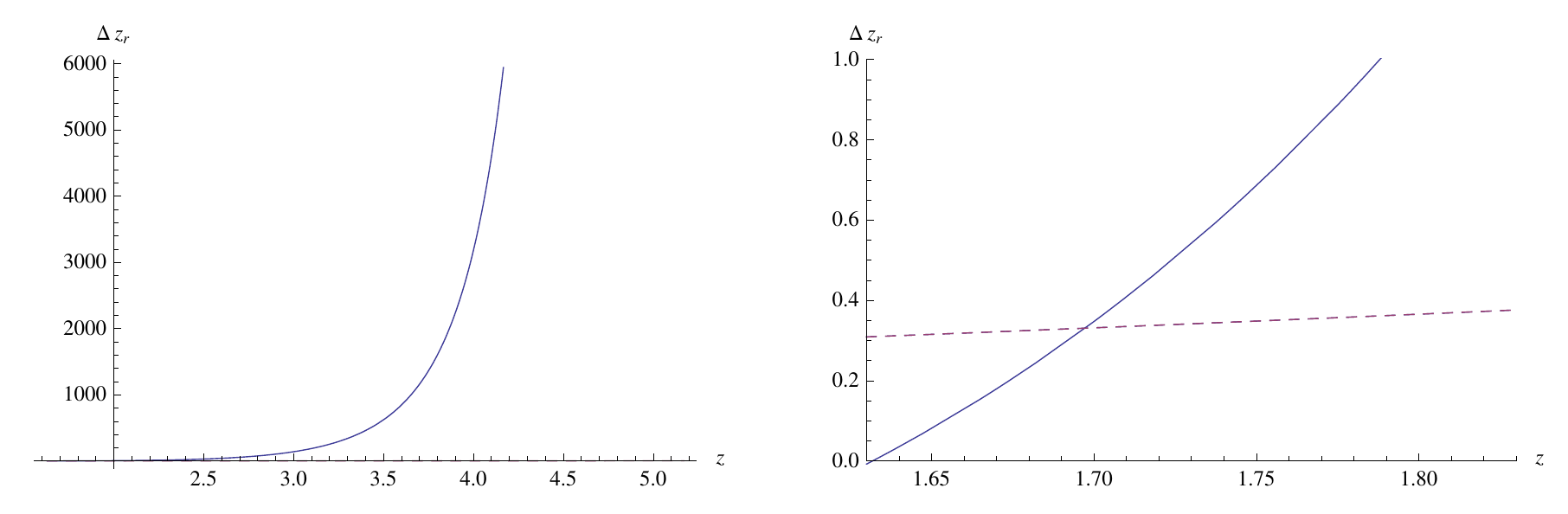} \caption{\footnotesize Both functions, $g_{\Lambda m}$ and  $z_r^{oi}$, can 
be interpreted as a difference between two $z_r$ values, and are presented as such in a $(\Delta z_r,z)$-plane. In both 
planes the solid curve is $g_{\Lambda m}$ and the dashed curve is $z_r^{oi}$. On the left hand 
side are $g_{\Lambda m}$ and $z_r^{oi}$ drawn throughout the values of $z$ where the nullcline can be attractive like, and on the right hand side are the same curves drawn at $1.63<z<2.00$. On the left hand side plot the scale of $\Delta z_r$ is such, that in practice the dashed curve is indistinguishable from the $z$-axis. From the right hand side plot one can see that at what range of $z$ values is $\delta_y/H_0$ smaller than observational inaccuracies.} 
\label{kuva_DE_matter_stab}
\end{figure}

\subsection{Dark energy and dust filled universe}

Next we assume universe to be homogeneous, flat and consists of dust and dark energy, thus from 
Eq. (\ref{observations2}) we obtain 
\be {observations_DE_matter}
z_r=H_0 (z+1) \sqrt{(z+1)^3 \Omega_m+\Omega_{ \Lambda }}.
\ee
Quantity $R(z)$ takes now a very inelegant form, which also is the case in every function including $R(z)$, 
hence none of the functions including $R(z)$ is explicitly presented in this subsection. Also, $R(z)$ includes 
an elliptic integral of the first kind, which makes accurate algebraic analysis difficult. However, numerical 
methods are sufficient enough in this case. 

Observations restrict the parameter values of this model: six-parameter $\Lambda$CDM model fit to WMAP nine-year 
data gives $\Omega_{\Lambda} = 0.721 \pm 0.025$, $\Omega_m =0.279 \pm 0.025 $, and $H_0 = 0.700 \pm 0.022$ \cite{nasa}. 
From Eq. (\ref{observations_DE_matter}) can be seen, that the effect of dark energy with $\Omega_m =0.279$, 
$\Omega_{\Lambda} = 0.721$ is less than one percent when $z \geq 5.37$, which is less than the inaccuracies of 
observed $\Omega_{m} $, $\Omega_{\Lambda} $, and $H_0$ values. Hence, at  $5.37 \leq z$ we approximate this model model to have 
$\Omega_{\Lambda}=0$. This approximation and numerical analysis at $0\leq z\leq 5.37$ reveals, that $L(z) = 0$ 
only at $z \approx 1.63$, and L and K are negative and J is positive approximately at $1.63 < z < 5.18$. 
In this case the boundary condition (\ref{eq3h}) reduces to be of the form
\be{ineq_DE_m}
\frac{\delta_y}{H_0}> g_{\Lambda m}(z;\Omega_{\Lambda},\Omega_m),
\ee
where the subscript $\Lambda m$ marks that this is the function $g(z;\Omega_i⁰)$ in the dust and dark energy filled case. The explicit form of the function $g_{\Lambda m}(z;\Omega_{\Lambda},\Omega_m)$ is neither elegant nor necessary to show here, thus it is not presented here. At $1.63 < z < 5.18$ the function $g_{\Lambda m}(z)$ is  monotonically 
increasing, and it is increasing in a very rapid fashion (as can be seen   from  Fig. \ref{kuva_DE_matter_stab}). 
In \cite{nasa} are the values of $\Omega_{\Lambda}$ and $\Omega_{m}$ given with error margin $\pm 0.025$, so function \be{oi}
z_r^{oi}:= \frac{|z_r(z;\Omega_{\Lambda}=0.696;\Omega_{m}=0.304)- z_r(z;\Omega_{\Lambda}=0.746;\Omega_{m}=0.254)|}{H_0}
\ee
represents the observational inaccuracy independent of parameter $H_0$. From the portrait on the right hand side of 
Fig. \ref{kuva_DE_matter_stab} can be seen, that $ g_{\Lambda m}(z)<z_r^{oi}$ at $1.63<z<1.70$, which therefore is the gap 
where this model can be attractive like. 

The time period corresponding $z$ values $1.63-1.70$ takes place about $9.70-9.85$ Gyr ago. Even though about $0.15$ 
Gyrs could be enough time for universe to evolve into similar state everywhere, thus possibly explaining the observations 
done at some range of $z$ values, it certainly can not explain the homogeneity of the observations at larger redshift values, 
{\it e.g.} over 10 Gyr old galaxies or CMB. The system with values $\Omega_m =0.279$, $\Omega_{\Lambda} = 0.721$, and 
$H_0 = 0.700$ is illustrated on the right hand side on Fig. \ref{kuva_molemmat_faasi}. The phase portrait looks similar 
to that of the dust filled case.

\section{Conclusions and discussion}

Phase space analysis is used to find out if there is isotropic dust source models describing the universe without fine 
tuning the initial state. In general with inhomogeneous models, it is not always reasonable to use term initial state  to refer only to the state of the system at Big Bang time, but sometimes also later times. This is due to observations: there is no observations of each location at all time during the entire history of the universe. The Big Bang might of occurred at different times in different regions, and the regions  might of evolved completely differently compared to each other. Hence, here the term is not fixed to some particular time or place, but is rather determined for each case separately, whatever is most convenient for the situation. Often we use it to refer to the time relatively near in the past of each observable, but there are some exceptions, such as homogeneous case, which are discussed more later.

When cosmological observations are plotted on the $(z_r,z)$-plane, the curve that is the best fit to those plots is the curve we refer as the best fit curve. An cosmological observable to be observed in the vicinity of the best fit curve today, does not demand that the observable have to evolve towards the curve 
throughout its history; it only needs to be there when observed. A single observable could in principle move 
repeatedly towards and away from the curve during its entire history, as long as it locates close enough the curve 
when we observe it. Though the amount of observables basically eliminates this kind of excessive behavior, but 
a reasonable situation would be where each observable at some point approach the best fit curve and stay close 
enough before observed.

The interpretation of the dynamical equations simplifies in the case of homogeneous models, as the initial state is 
same everywhere and the Big Bang time occurred simultaneous everywhere; none of the quantities in equations governing the evolution are dependent of $r$, and the evolution have been similar always and everywhere. 
Thus, the minimum requirement for a FLRW metric based on model to be viable and stable enough is, that there has been a period  
where it have evolved towards the best fit curve and located inside boundaries given by observation inaccuracies since. 
This is the minimum requirement in the sense, 
that there was at least one early period where the best fit curve  was attractive. 
The given example of the dust and dark matter filled case have an era where redshift is approaching the best fit 
curve: about 9.70-9.85 Gyrs ago. Hence the model could explain almost the last 10 Gyrs. \footnote{Since we do not have a 
general method to determine if an observable locates close enough the best fit curve when it is not attractive, 
we can not say if it really is so. So far the only way to determine this is to examine each case separately.}
Thus the model can not explain oldest observations. This is not 
surprising: as it is well known, without inflation homogeneous models have a fine tuning problem of matter and energy density.\cite{KolbTurner1994, Weinberg2008}  
In the light of this paper, the stability is not a inbuilt feature at least in the cases studied in this 
paper, and can not replace the widely excepted solution of early time inflation for the fine tuning problem.

Analogous results are expected for almost homogeneous cases, where initial state does not vary too much with respect to $r$ 
and the evolution can be though of almost similar everywhere: observations can be explained without fine tuning the 
initial values, if at some point in the history before CMB there was a period where the universe evolved towards the 
best fit curve (and have been close enough it since). Thus we can conclude, that if the observed universe have not 
had a period in its history before CMB where the boundary condition we gave is satisfied, LT model can not explain its homogeneous nature without fine tuning.

The redshift values where the system is not defined, singular limits, causes difficulties for analysis. In this paper 
singular limits are not studied, and it is left to forthcoming publications. Nevertheless, one note of singular limits should be 
brought out here.
In fig. \ref{kuva_molemmat_faasi} is illustrated what happens close by $z=5/4$, where the system is not defined. Even a slightest
variation from today's observed value of $H_0$ seams to lead a solution into very different direction at the singular 
limit. On the other hand, if we would of drawn the phase plane with {\it e.g.} $H_0=0.71$, the solution with $H_0=0.71$ 
would of approached towards point $\sim (1.25,5)$ and the solutions with $H_0=0.70$ and $H_0=0.69$ would of 
approached towards the $z$-axis. This may be a sign, for example, of an instability of the model or of a break 
down of the used method at this $z$ value and its close neighborhood. As singular limits in general, also this 
phenomena is planned to be investigated thoroughly later.

The overall view our analysis cast on inhomogeneous models is not very promising. The best fit curve that should 
be attractive like is approximately given with Eq. (\ref{observations_DE_matter}), thus increasing at rate $z^{5/2}$ with large $z$ 
values, and rapid increase of the curve makes it more unlikely to be attractive like. The situation seems to get 
worse if pressure is taken into account, since the increasing rate is then proportional to $z^3$ with large redshift. 
However, it is possible for observationally acceptable inhomogeneous (and therefore also homogeneous) models to be 
attractive like even with large $z$ values, hence it needs to be investigated. The situation, however, is not 
necessarily as bad as it looks, since even though pressure at first seems to make the situation worse, it may 
actually recover it. This is due to the chances pressure brings with it to the evolution equation. For example, 
if the evolution equation changes from second order differential equation to higher order one, it may chance the
structure of the stability of the system dramatically. This may even happen for homogeneous models, thus the result 
here received for dust and dark energy filled universe should not be interpreted as final.



\end{document}